\documentclass[12pt,epsfig]{article}
\usepackage{epsfig,graphicx,amsmath,amssymb}

\newlength{\dinwidth}
\newlength{\dinmargin}
\def\lapproxeq{\lower .7ex\hbox{$\;\stackrel{\textstyle
<}{\sim}\;$}}
\def\gapproxeq{\lower .7ex\hbox{$\;\stackrel{\textstyle
>}{\sim}\;$}}
\def\gtrsim{\lower .7ex\hbox{$\;\stackrel{\textstyle
>}{\sim}\;$}}
\def\lesim{\lower .7ex\hbox{$\;\stackrel{\textstyle
<}{\sim}\;$}}

\def\be{\begin{equation}}
\def\ee{\end{equation}}
\def\bea{\begin{eqnarray}}
\def\eea{\end{eqnarray}}

\def\bb{b\bar{b}}

\def\ra{ \rightarrow }

\def\ra{\rightarrow}
\def\lapproxeq{\lower .7ex\hbox{$\;\stackrel{\textstyle
<}{\sim}\;$}}
\def\gapproxeq{\lower .7ex\hbox{$\;\stackrel{\textstyle
>}{\sim}\;$}}
\def\bb{b\bar{b}}

\def\be{\begin{equation}}
\def\ee{\end{equation}}
\def\bea{\begin{eqnarray}}
\def\eea{\end{eqnarray}}

\begin{document}
%\titlepage
\begin{flushright}
\today \\

\end{flushright}

\vspace*{0.5cm}
\begin{center}
{\Large{\bf Diffractive Higgs production}}

\vspace{5mm}
\textsc{{A.D.~Martin}$^a$, V.A. Khoze$^a$ and M.G. Ryskin$^{a,b}$}

\vspace{2mm}
{$^a$IPPP, Physics Department, University of Durham, DH1 3LE, UK\\
$^b$Petersburg Nuclear Physics Institute, Gatchina, St. Petersburg, 188300, Russia}
\end{center}

\vspace{3mm}
\begin{abstract}
We review the advantages of observing exclusive diffractive Higgs production at the LHC. We note the importance of the Sudakov form factor in predicting the event rate. We discuss briefly other processes which may be used as `standard candles'. 
\end{abstract}

\section{Introduction}
Central exclusive diffractive (CED) processes offer an excellent opportunity to study the Higgs sector at the LHC in an exceptionally clean environment. The process we have in mind is
\begin{equation}
 \label{excl}
 pp\to p\; +\; H\; +\; p
\end{equation}
where the + signs denote large rapidity gaps.  We consider the mass range, $M \lapproxeq 140$ GeV, where
the dominant decay mode is $H \to \bb$. 
Demanding such an exclusive process (\ref{excl}) leads to a small cross section \cite{KMRplb}.
At the LHC, we predict
\begin{equation}
\sigma_{\rm excl}(H)~\sim ~10^{-4}~\sigma^{\rm tot}_{\rm incl}(H).
\end{equation}
In spite of this, the exclusive reaction (\ref{excl}) has the following advantages:
\begin{itemize}
\item[(a)]
The mass of the Higgs boson (and in some case the width) can be measured
with high accuracy (with mass resolution $\sigma(M)\sim 1$ GeV) by measuring the
missing mass to the forward outgoing protons, {\it provided} that they can be accurately tagged some 400 m from the interaction point.
\item[(b)]
The leading order $b\bar b$
  QCD background is suppressed by the P-even $J_z=0$ selection
rule \cite{KMRmm}, where the $z$ axis is along the direction of the proton beam.
Therefore one can observe the Higgs boson via the main decay mode $H\to
b\bar b$. Moreover, a measurement of the mass of the decay products must match the `missing mass' measurement. It should be possible to achieve a signal-to-background ratio of the order of 1. For an integrated
LHC luminosity of ${\cal L} =300 ~{\rm fb}^{-1}$ we predict about 100 observable Higgs events, {\it after} acceptance cuts \cite{DKMOR}; assuming pile-up problems have been overcome.
\item[(c)]
The quantum numbers of the central object (in particular, the
C- and P-parities) can be analysed by studying the azimuthal angle
distribution of the tagged protons \cite{Centr}. Due to the selection
rules, the production of $0^{++}$ states is strongly favoured.
\item[(d)]
There is a very clean environment for the
exclusive process -- the soft background is strongly suppressed.
\item[(e)]
Extending the study to SUSY Higgs bosons, there are regions of SUSY parameter space were the
signal is enhanced by a factor of 10 or more, while the background remains unaltered.  Indeed,
there are regions where the conventional inclusive Higgs processes are suppressed and the CED signal is
enhanced, and even such that both the $h$ and $H$ $0^{++}$ bosons may be detected \cite{KKMRext}.   
\end{itemize}

\section{The cross section: the role of the Sudakov form factor}

\begin{figure}
\begin{center}
\epsfig{file=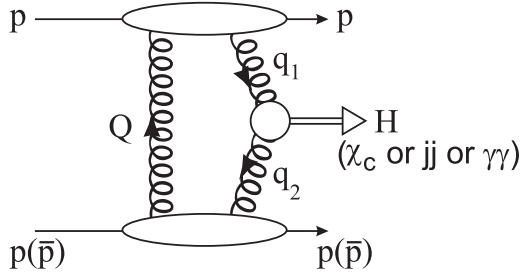,width=0.4\textwidth}
%\centerline{\epsfxsize=0.4\textwidth\epsfbox{slide.eps}}
\caption{Schematic diagram for central exclusive  production,
$pp \to p+X+p$. The presence of Sudakov form factors ensures the infrared
stability of the $Q_t$ integral over the gluon loop. It is also necessary
to compute the probability, ${\hat S}^2$, that the rapidity gaps survive soft rescattering.}
\label{fig:H}
\end{center}
\end{figure}

The basic mechanism for the exclusive process, $pp\ra p+H+p$, is
shown in Fig.~$\ref{fig:H}$. The left-hand gluon $Q$ is needed to screen the
colour flow caused by the active gluons $q_1$ and $q_2$. The cross section is of the form \cite{KMR,KMRmm}
\begin{equation}
\sigma \simeq {\hat S}^2 \left| N\int\frac{dQ^2_t}{Q^4_t}\: f_g(x_1, x_1', Q_t^2, \mu^2)f_g(x_2,x_2',Q_t^2,\mu^2)~ \right| ^2, \label{eq:M}
\end{equation}
where the constant $N$ is known in terms of the $H\to gg$ decay width \cite{INC,KMR}. 
The first factor, ${\hat S}^2 $, is the probablity that the rapidity gaps survive against population by secondary hadrons from the underlying event, that is hadrons originating from {\it soft} rescattering. It is calculated using a model which embodies all the main features of soft diffraction.  It is found to be ${\hat S}^2 =0.026$ for $pp\ra p+H+p$ at the LHC. 
The remaining factor, $|...|^2$, however, may
be calculated using perturbative QCD techniques, since the dominant contribution to the integral comes from the region $\Lambda_{\rm QCD}^2\ll Q_t^2\ll M_H^2$. 
The probability amplitudes, $f_g$, to find the appropriate pairs of
$t$-channel gluons ($Q,q_1$) and ($Q,q_2$), are given by the skewed
  unintegrated gluon densities at a {\it hard} scale $\mu \sim M_H/2$.

Since the momentum fraction $x'$ transfered through the
screening gluon $Q$ is much smaller than that ($x$) transfered through
the active gluons $(x'\sim Q_t/\sqrt s\ll x\sim M_H/\sqrt s\ll 1)$, it
is possible to express $f_g(x,x',Q_t^2,\mu^2)$
in terms of the conventional integrated density
$g(x)$. A simplified form of this relation is \cite{KMR}
\begin{equation}
\label{eq:a61}
  f_g (x, x^\prime, Q_t^2, \mu^2) \; = \; R_g \:
\frac{\partial}{\partial \ln Q_t^2}\left [ \sqrt{T_g (Q_t, \mu)} \: xg
  (x, Q_t^2) \right ], 
\end{equation} 
which holds to 10--20\%
accuracy.
The factor $R_g$ accounts for
the single $\log Q^2$ skewed effect.  It is found to
be about 1.4 at the Tevatron energy and about 1.2 at the energy of the LHC.

Note that the $f_g$'s embody a Sudakov suppression
factor $T$, which ensures that the gluon does not radiate in the
evolution from $Q_t$ up to the hard scale $\mu \sim M_H/2$, and so
preserves the rapidity gaps. The Sudakov factor is \cite{WMR}
\begin{equation}
\label{eq:a71}
  T_g (Q_t, \mu)=\exp \left (-\int_{Q_t^2}^{\mu^2}
  \frac{\alpha_S (k_t^2)}{2 \pi}\frac{dk_t^2}{k_t^2} \left[
%  \int_0^1 \: \left [\Theta(1-z-\Delta)\Theta(z-\Delta)zP_{gg} (z) \:
  \int_\Delta^{1-\Delta}zP_{gg} (z)dz
\ + \ \int_0^1 \sum_q\
  P_{qg} (z)dz\right]\right),
\end{equation}
with $\Delta = k_t/(\mu + k_t)$.  The square root arises in
(\ref{eq:a61}) because the (survival) probability not to emit any
additional gluons  is only relevant to
the hard (active) gluon.  It is the presence of this Sudakov factor
which makes the integration in (\ref{eq:M}) infrared stable, and
perturbative QCD applicable \footnote{Note also that the Sudakov factor inside the loop
integration induces an additional strong decrease (as $M^{-3.3}$ for $M \sim 120$ GeV) of the cross section as the mass $M$ of the centrally
produced hard system increases \cite{KKMRext}.  Therefore, the price to pay for
neglecting this suppression effect would be to considerably
overestimate the central exclusive cross section at large masses.}.

It should be emphasised that the presence of the double
logarithmic $T$-factors is a purely classical effect, which  was first
discussed in 1956 by Sudakov in QED.  There is strong brems-strahlung
when two colour charged gluons `annihilate' into a heavy neutral object and the
probability not to observe such a bremsstrahlung is given by the
Sudakov form factor \footnote{It is worth mentioning that the $H\to gg$ width entering
the normalization factor $N$ in (\ref{eq:M}) is an
`inclusive' quantity which includes all possible bremsstrahlung
processes. To be precise, it is the sum of the $H\to gg+ng$ widths, with
$n$=0,1,2,... . The probability of a `purely exclusive' decay into two
gluons is nullified by the same Sudakov suppression.}.
  Therefore, any model (with perturbative or
non-perturbative gluons) must account for the Sudakov suppression when
producing exclusively a heavy neutral boson via the fusion of two
coloured/charged particles.

In fact, the
$T$-factors can be calculated to {\it single} log
accuracy \cite{KKMRext}. The collinear single logarithms may be summed up using the
DGLAP equation. To account for the `soft' logarithms (corresponding
to the emission of low energy gluons) the one-loop virtual correction
to the $gg\to H$ vertex was calculated explicitly, and then the scale
$\mu=0.62\ M_H$ was chosen in such a way that eq.(\ref{eq:a71})
reproduces the result of this explicit calculation. It is sufficent to
calculate just the one-loop correction since it is known that the
effect of `soft' gluon emission exponentiates. Thus
(\ref{eq:a71}) gives the $T$-factor to single log accuracy.

In some sense, the $T$-factor may be considered as a `survival'
probability not to produce any hard gluons during the $gg\to H$ fusion
subprocess. However it is not just a number (i.e. a numerical factor) which
may be placed in front of the integral (the `bare amplitude'). Without the
$T$-factors hidden in the unintegrated gluon densities $f_g$ the integral
(\ref{eq:M}) diverges. From the formal viewpoint, the suppression of
the amplitude provided by $T$-factors is infinitely strong, and without them
the integral depends crucially on an ad hoc infrared cutoff.

\section{`Standard candles': calibrating the exclusive Higgs signal}

As discussed above, the exclusive Higgs signal is particularly clean, and the signal-to-background
ratio is favourable.
 However, the expected number of events in the SM case is low.
Therefore it is important to check the predictions for exclusive Higgs production
by studying processes mediated by the same mechanism, but
with rates which are sufficiently high, so that they may be observed at the Tevatron
(as well as at the LHC).  The most obvious examples are those in which the Higgs 
is replaced by either a dijet system, a $\chi_c$ or $\chi_b$ meson, or
by a $\gamma \gamma$ pair, see Fig.~$\ref{fig:H}$. 

CDF have made a start. They have a value for exclusive $\chi_c$ production; after acceptance cuts they find \cite{CDF}
$\sigma(\chi_c \to \mu\mu\gamma) \sim 50$ pb, with a large uncertainty. This happens to be equal to the KMR prediction \cite{KKMRS} for the same cuts, which, because of the low scale, is only an order-of-magnitude estimate.  Exclusive $\gamma\gamma$ production is a clean signal, but the rate is quite low \cite{KMRS}.

Here, therefore, we discuss the exclusive production
of a pair of high $E_T$ jets, $p\bar {p} \to p+jj+\bar {p}$. 
 The corresponding cross section \cite{KMR,INC} was evaluated to
be about 10$^4$ times larger than that for the SM Higgs boson.
Thus, in principle,
this  process appears to be an ideal `standard candle'.  The expected cross section is rather large,
and we can study its behaviour as a function of the mass of the dijet
system.  This process is being studied by the CDF collaboration. Unfortunately, in the present CDF environment, the separation
of exclusive events is not unambiguous.
At first sight, we might expect that the exclusive dijets form a narrow peak,
sitting well above the background, in the
distribution of the ratio
\begin{equation}
R_{jj}=M_{{\rm dijet}}/M_{\rm {PP}}
\end{equation}
at $R_{jj}=1$, where $M_{\rm {PP}}$ is the invariant energy of the incoming
 Pomeron-Pomeron system.  In reality
the peak is smeared out due to hadronization, the jet-searching algorithm and detector effects.
Moreover, since $M_{{\rm dijet}}$ is obtained from measuring just the two-jet part of the exclusive signal; there will be a `radiative tail' extending to lower values of $R_{jj}$.

The estimates \cite{KKMRS} give an exclusive cross section for dijet
production with $E_T> 10,  25,  35, 50$ GeV, with values which are comparable to the recent CDF values \cite{CDF}, based on events with $R_{jj} > 0.8$.  
As discussed above,
one should not expect a clearly `visible' peak in the CDF data for $R_{jj}$ close to 1.
 It is worth mentioning that the CDF measurements have already started to reach values of the invariant mass of the Pomeron-Pomeron system in the SM Higgs mass range.

\section*{Acknowledgments}
ADM thanks the Leverhulme Trust for an Emeritus Fellowship and the Royal Society for a Joint Project Grant with the former Soviet Union.  

%\section*{References}

\end{document}